\documentclass{article}
\usepackage{graphicx} 

\title{Wigner's Frame}
 \author{Emily Adlam \thanks{Philosophy Department and Institute for Quantum Studies, Chapman University, Orange, CA92866, USA \texttt{eadlam90@gmail.com} }}
\date{\today}
\usepackage{amsmath}
\usepackage{natbib} 

\begin{document}

\maketitle

\begin{abstract}
    This article suggests that thinking about the role of reference frames can provide new insight into   Extended Wigner's Friend scenarios. This involves appealing to symmetries to make a principled distinction between properties of a system which are meaningful only relative to an external reference system and properties which are meaningful without further relativization. Thus we may propose that there are always well-defined facts about what observers have observed, but there are not necessarily well-defined facts about the relations between their reference frames, so there will not always exist a  joint distribution over their outcomes which can meaningfully be compared to the predictions of quantum mechanics. In addition, this approach also offers a general argument against the idea that there should be a regress of relativization. 
\end{abstract}

\section{Introduction}

Recent years have seen significant developments in the study of quantum reference frames and relational descriptions. And because there is surely some connection between the role of reference frames and the role of observers, it is natural to think  this line of research should have something to tell us about Wigner's Friend scenarios and possibly the measurement problem. In this article I will explore some of these  connections.

Many people have the intuition that the  Wigner's Friend scenario \citep{Wigner} should be resolved by positing some form of indefiniteness. In recent years, various Extended Wigner's Friend scenarios \citep{schmid2023review} have been proposed to force an explicit contradiction between various natural assumptions and the predictions of quantum mechanics, and it has  been suggested that we should resolve the apparent contradiction by denying the `absoluteness of observed events.'  This involves supposing that there is no absolute fact of the matter about what a given observer has observed, and thus for a set of outcomes observed by different observers there simply does not exist a well-defined probability distribution over these outcomes which can be compared to the predictions of quantum mechanics.

But  denying absoluteness in this sense leads to significant epistemic and methodological difficulties, so it would perhaps be preferable if we could find a different way to use indefiniteness to resolve the apparent contradiction. And a natural possibility is to connect this issue to reference frames. For the result of a measurement performed on a quantum degree of freedom is defined relative to a reference system, so for example the outcome of a spin measurement is stated relative to a rod which defines an axis; but the facts about what has been observed in such a measurement, e.g. `I found the spin of the particle to be parallel to the axis,' do not obviously need to be relativized to any further reference system. Leveraging this distinction may  offer an appealing way of addressing the Wigner's Friend conundrum.

For example, suppose we postulate that in such an experiment it is possible for the relation that a laboratory system bears to external reference systems to be indefinite, but the properties of the integrated laboratory system which are meaningful in the absence of an external reference frame are never indefinite. So when a measurement is performed on a quantum degree of freedom, there is always a well-defined absolute fact of the matter about what the person who performed it observed, relative to their own reference frame, but there is not necessarily a fact of the matter about the relationships between the frames of reference employed by different observers - for example, there is not necessarily a well-defined angle between the axes used by different observers. This would imply that for any set of measurements there is always a well-defined joint probability distribution over the actual \emph{observations} that are made, but such a distribution does not necessarily correspond to a well-defined joint probability distribution over a set of variables all defined relative to the same reference frame. Thus since quantum mechanics only makes joint predictions for outcomes defined relative to a common frame of reference, the absoluteness of the observations does not imply the existence of a well-defined joint distribution which can meaningfully be compared to the predictions of quantum mechanics, so we can avoid the contradiction without denying the absoluteness of observed events. 
 
  Moreover, I will suggest here that such an approach follows  naturally from an examination of the symmetries of these kinds of systems. This is because the properties which are commonly associated with quantum degrees of freedom, such as spin orientation, position, energy and so on, are the types of properties which can be altered by transformations which are symmetries of the relevant subsystems, and thus such properties are meaningful only relative to an external reference system; whereas facts about what has been observed, such as `I found the spin of the particle to be parallel to the axis'   plausibly live in the invariant subspace of the relevant subsystems with respect to  symmetry transformations, and therefore such properties do not need to be relativized to an external reference frame. This offers a helpful way of making a principled distinction which can be applied to the Wigner's Friend scenarios.

  I will begin in section \ref{Wigner} by introducing the Bong et al Extended Wigner's Friend scenario and suggesting a way of resolving the paradox by leveraging the distinction between external relational properties and internal invariant properties. Then in section \ref{symmetry} I will elaborate on the connection with physical symmetries. In section \ref{regress} I will discuss a possible objection associated with the idea that there is a regress of relativization, and finally in section \ref{intersubjectivity} I will discuss how this idea is related to some existing approaches to the measurement problem, particularly Healey's ideas about decoherence environments.

\section{Wigner's Friend \label{Wigner}}
 
In the basic Wigner's Friend scenario \citep{Wigner}, we imagine some observer, Alice, equipped with a qubit $S$  in the state $a | \downarrow \rangle + b |\uparrow \rangle$, with both $a$ and $b$ non-zero. Suppose Alice performs a spin measurement with possible outcomes `down' ($\downarrow$) and `up' ($\uparrow$) associated with the operators $\{ |\downarrow \rangle \langle \downarrow |, |\uparrow \rangle \langle \uparrow | \}$.  It is natural to suppose that Alice will see an  outcome to her measurement,  and   quantum mechanics as standardly applied tells us that the system is now in the state corresponding to that outcome:   if Alice sees the outcome `down', then the state of system $S$ is now $| \downarrow \rangle$, and if Alice sees the outcome `up' then the  state of system $S$ is now $|\uparrow \rangle$. 

However, suppose that Bob, observing from the outside, is able to maintain Alice's laboratory system in a coherent state, and he describes the whole interaction using unitary quantum mechanics. He will conclude that  $S$ and Alice are now in the following entangled superposition state: 

\[ \psi_{SA} = a | \downarrow \rangle_S | \downarrow \rangle_A + b |\uparrow \rangle_S |\uparrow \rangle_A \]

Here, $|\downarrow\rangle_A$ and $| \uparrow\rangle_A$ are states of Alice corresponding to `seeing the down outcome' and `seeing the up outcome' respectively. 

Something strange seems to be happening here: surely Alice  must have  experienced a single outcome, but the state assigned by Bob appears to represent Alice as somehow experiencing a combination of two different outcomes. Moreover, if we think quantum mechanics is universally correct in its predictions, we cannot say that the state relative to Bob just describes  his lack of knowledge of the true outcome; for the universality of quantum mechanics implies that if Bob measures Alice and $S$ he will  see outcomes corresponding to the entangled state $\psi_{SA}$ rather than to either $|\downarrow\rangle$ or $| \uparrow \rangle$, so in some robust sense  $\psi_{SA}$ really is the correct state relative to him.

The worry here can be sharpened by considering an `Extended Wigner's Friend' scenario which combines two Wigner's Friend experiments with a Bell experiment. There are various theorems of this kind, but let us take as an example the Local Friendliness theorem due to \cite{Bong_2020}. Here we have Alice and Chidi perform measurements on two halves of an entangled two-qubit state, obtaining outcomes $A, C$, and then Bob keeps Alice's laboratory in a coherent state and performs on a certain kind of measurement which does not commute with Alice's original measurement, obtaining an outcome $B$, while Divya does the same with Chidi, obtaining an outcome $D$. Then it can be shown that for certain choices of measurements for the four observers, there is no probability distribution over the four measurement outcomes $A, B, C, D$ which reproduces all the pairwise predictions of quantum mechanics for the pairs $\{ (B,D), (B, C), (A, D), (A, C) \}$ while also satisfying a locality assumption plus a no-retrocausality and no-superdeterminism assumption. 

Now, it should be noted that Bob necessarily erases Alice's outcome $A$ when performing the measurement to find out the value of $B$, and likewise Divya erases $C$ when performing a measurement to find out the value of $D$. So initially it looks as though the only pair of measurements that anyone can actually compare is $(B, D)$. Thus one might try to avoid the paradox by insisting that the predictions of quantum mechanics can only be applied in a \emph{first-person} sense \citep{adlamnonabsolute}: that is, quantum mechanics always makes the correct predictions for anything which can be observed by an individual observer, but its predictions are not  necessarily applicable for pairs of outcomes which can never be compared. In that case it would seem that only the pair $(B,D)$ needs to obey the predictions of quantum mechanics, so no contradiction arises. 

But we can imagine giving Bob a choice. Either  he performs the measurement to acquire the value $B$, or else he performs a `measurement' in a basis which is suited to establish the result of Alice's measurement, in which case he acquires the value $A$ instead (for example, perhaps he simply talks to Alice and asks her about her outcome). Divya can  be given a similar option to choose between acquiring the value $D$ or acquiring the value $C$ instead. This suggests that although only one of the pairs $\{ (B,D), (B, C), (A, D), (A, C) \}$ will be observed on any given run, any one of these pairs \emph{could} be observed on any given run. So if we assume that quantum mechanics always makes the correct predictions  at least in a first-person sense, it follows that statistically, over the course of many repetitions of the experiment, the  pair  actually observed must obey the quantum predictions. Moreover if we rule out  retrocausality and superdeterminism, then the outcomes obtained by Alice and Chidi can't depend on the later choices by Bob and Divya, and if we rule out non-locality then Bob's outcome can't depend on Divya's choice or vice versa, so it seems that there must be a single fixed probability distribution over the four variables $A, B, C, D$ which reproduces  the pairwise predictions of quantum mechanics for \emph{all} the pairs $\{ (B,D), (B, C), (A, D), (A, C) \}$.   Thus even if we only insist on the correctness of quantum mechanics in a first-person sense,  we still end up with a contradiction. 

One popular proposal for responding to this paradox involves  denying the `absoluteness of observed events,' which is the assumption that observed events are absolute single events, not relative to anything. If observed events are not absolute in this sense then there is no fact of the matter about the relationships between the observations made by different observers, and so there  cannot exist any well-defined probability distribution over the four outcomes $A, B, C, D$ which could be compared to the predictions of quantum mechanics.  Thus, it is suggested, we can maintain all the predictions of quantum mechanics without allowing non-locality, retrocausality or superdeterminism. 

However, on the face of it non-absoluteness of this kind seems very problematic from the point of view of the epistemology of science \citep{https://doi.org/10.48550/arxiv.2203.16278,adlamnonabsolute}, because it seems to imply that there is some kind of instability in memory, communication or language itself, leading to a picture in which each of us is trapped within a set of facts relativized to the self and/or to the present moment, with no possible means of learning about any regularities existing outside of this one set of relative facts. Of course, this difficulty could perhaps be mitigated by putting forward some principled reason why  failures of memory, communication or language occur only in \emph{this} specific kind of case. But the worry here is that in the scenario where Bob simply has a conversation with Alice to acquire the value of $A$, there does not seem to be anything particularly special about the way in which memory, communication and language are employed, so it's unclear that there is any sufficiently robust justification which could reassure us that the epistemic difficulties that arise here will remain limited to the specific EWF scenario. So this route remains an option, but it would need to be handled with care.

\subsection{Reference Frames and Wigner's Friend}

For these reasons, I think that approaches which deny `the absoluteness of observed outcomes' should probably be rejected. But there may be a way in which we can appeal to a kind of indefiniteness in the relation between Alice and Bob whilst nonetheless maintaining absoluteness of observed outcomes in the epistemically relevant sense. 

The idea here is to  recall that properties such as position, spin, energy and so on are meaningful only relative to some choice of reference frame, and then note that in the EWF experiment the four observers are all using different physical systems to define their reference frames, so their measurement outcomes cannot be stated relative to a common reference frame unless there is some well-defined relation between their individual reference frames. Thus, since quantum mechanics only makes predictions for outcomes stated relative to a common reference frame, it follows that even if  there are always well-defined `absolute' facts about what each observer has observed relative to their own reference frame, it may not necessarily be possible to compare the probability distribution over these facts to any quantum predictions. 

To be concrete, I will suppose henceforth that as described above, Alice is making a measurement of the spin of a particle $S$ by comparing the direction of the spin to a fixed physical axis that she stores in her laboratory - in section \ref{symmetry} I will discuss the case where Alice measures some other kind of variable instead. At the time of the measurement Alice makes some observation;  that is, either she sees `the spin of $S$ is parallel to my axis,' or she sees `the  spin of $S$ is antiparallel to my axis.' Let us suppose that on every run of the experiment there is always a well-defined absolute fact about whether Alice saw parallel or antiparallel. However, suppose also that immediately after Alice's measurement there is not in general a well-defined fact of the matter about the orientation of Alice's axis relative to Bob's axis. So  even though there is a well-defined fact about what Alice observed, there is not necessarily a well-defined fact about the orientation of the spin of $S$ relative to Bob's reference frame. 

Furthermore, let us suppose here that Alice's reference frame acquires a definite orientation relative to Bob's reference frame only if and when Bob performs a measurement in a basis which is suited to establish the result of Alice's measurement - i.e. in the case where he chooses to acquire the value of $A$ rather than the value of $B$. Thus although there is always a well-defined absolute fact about what Alice has observed, there comes to be a fact about the orientation of the spin of $S$ relative to Bob's reference frame only on the runs where Bob chooses to acquire the value $A$. I will make the same suppositions for Chidi and Divya. In section \ref{symmetry} I will consider what kind of underlying physical story might render these suppositions plausible, but for now let us see how they help resolve the Wigner's Friend paradox. 

As noted, the proof of \cite{Bong_2020} proceeds by arguing that if there are `absolute' facts about what everyone sees, then we must be able to write down a well-defined probability distribution over the four measurement outcomes $A, B, C, D$. But there is an ambiguity in the definition of the variable $A$. It could be the variable defined by Alice's \emph{own internal observation},  $A_I$, which takes values $\{ -1, +1\}$ such that $-1$ means `Alice found the particle to be antiparallel  to her axis' and $+1$ means `Alice found the particle to be parallel  to her axis.' Or it could be the variable $A_E$ stating the direction of the spin \emph{relative to Bob's reference frame}, which takes values $\{ -1, +1 \}$ such that $-1$ means `the particle is antiparallel to Bob's axis,' which corresponds to either the case where Alice found the particle to be parallel to her axis and that axis is then found to point in the opposite direction to Bob's axis, or alternatively Alice found the particle to be antiparallel to her axis and that axis is then found to point in the same direction as Bob's axis' (and similarly mutatis mutandis for $+1$). The proposal above then amounts to the supposition that on every run of the experiment there is always a well-defined fact about the value of $A_I$, but there is not necessarily a well-defined fact about the value of $A_E$. 

Now when we talk about the `absoluteness of observed outcomes,' it seems natural to suppose that this phrase is intended to refer to variables like $A_I$, rather than $A_E$. This is because the point of emphasizing that the outcomes are `observed' is precisely to remind us that the facts at issue  pertain to what Alice \emph{actually experiences}, since  there is a strong intuition that experiences are intrinsic to the person to whom they belong and cannot be relativized to anything else. This motivation is already present in the original Bong et al paper and is particularly emphasized by \cite{Wiseman_2023}, where the  reasons for expecting some kind of `absoluteness' for  the values of $A$ and $C$ are based on principles like `ego absolutism,' -  `My communicable thoughts are absolutely real' - and `friendliness' -  `If a system displays independent cognitive ability at least on par with my own, then they are a party with cognition at least on par with my own, and any thought they communicate is as real as any communicable thought of my own.' So in this presentation it seems clear the `observed event' is the one encoded in the variable $A_I$ rather than the variable $A_E$, since $A_I$ is the one which correlates with the internal thoughts that Alice will actually be having when she makes the observation. In addition, when we are concerned with the kinds of epistemic worries discussed in the previous section, what matters is that Alice can remember her actual experience  and communicate it to Bob, so  $A_I$ is the relevant variable from the epistemic point of view. Thus we should say that  `absoluteness of observed outcomes' is satisfied provided that there exists a well-defined joint probability distribution over the four variables $A_I, B, C_I, D$, where $C_I$ is defined similarly to $A_I$\footnote{Technically we should probably also use $B_I$ and $D_I$ as well, but the difference between $B_I$ and $B_E$ etc isn't important as long as we don't bring in another external observer who keeps   Bob and his laboratory in a coherent state and performs a certain kind of measurement on it.}. 

 However, we cannot possibly compare the distribution over $A_I, B, C_I, D$ to the predictions of quantum mechanics for  a pair such as $(A_I, D)$. This is because quantum mechanics makes predictions for correlations between these pairs of variables only where both variables are stated \emph{relative to the same reference frame}, i.e. there must be a shared set of axes to which their orientation can be compared, which is not the case for $A_I$ and $D$. Of course as long as there is a well-defined relative orientation between Divya's frame and Alice's frame we can simply rotate the predictions appropriately and thus still arrive at a statement about  what the  correlations between $A_I$ and $D$ should be according to quantum mechanics. But if there is no fact of the matter about the relationship between Alice's axes and Divya's axes, then quantum mechanics will have nothing to say about what correlations we should expect to see between $A_I$ and $D$.

Thus once we switch to using $A_I$ rather than $A_E$, in order to be able to compare the distribution over $A_I, B, C_I, D$ to some  quantum predictions we are going to need  some extra variables. Assuming that in the case where Bob chooses to acquire the value $A$ the axis of Alice is always found to be either parallel or antiparallel to Bob's axis, let us define a variable $A_R$ which equals $-1$ if they are found antiparallel and $+1$ if they are found parallel; define $C_R$ similarly for Chidi. Then in order to be able to make a comparison to the  predictions of quantum mechanics we will need to write down a distribution over the \emph{six} variables $A_I, B, C_I, D, A_R, A_R$, where $A_I$ and $A_R$ will subsequently be combined to arrive at the variable $A_E$ which records the direction of Alice's spin relative to the reference frame shared by Bob and Divya, and similarly for $C_I$ and $C_R$; that is, $A_E = A_I \times A_R$, and   $C_E = C_I \times C_R$.\footnote{More generally we might imagine that it is possible for Alice's axis to be found pointing in a wider variety of  directions relative to Bob's axis, or indeed perhaps it could point in any direction. If so, we can simply generalize this description by defining a variable $\vec{a}$ giving the direction in which Alice's axis points relative to Bob's axis; then we would define $\vec{A}_E = A_I \times \vec{a}$, so we end up with some vector quantity specifying the direction of spin of Alice's particle relative to Bob's reference frame, so it  can now be compared to the quantum predictions.} 

However, here we have adopted the supposition that Alice's axis comes to have a well-defined orientation relative to Bob's axes only in the case where Bob does not measure $B$ and instead acquires $A$. This means that in fact $B$ and $A_R$ will never concurrently exist; if Bob measures $B$, then he cannot also bring Alice's axis into alignment with his, so $A_R$ is not well-defined and hence $A_E$ is not well-defined either. The same applies to $D$ and $C_R$ and $C_E$. So in this setting, although there is a well-defined probability distribution over  $A_I, B, C_I, D$, there is \emph{not} a well-defined probability distribution over $A_I, B, C_I, D, A_R, A_R$, and hence there is also not a well-defined probability distribution over $A_E, B, C_E, D$. Thus once we account for the role of reference frames  the non-absoluteness theorem cannot be formulated, because  absoluteness in the epistemically relevant sense (the existence of a well-defined distribution over $A_I, B, C_I, D$) does not imply the existence of a joint probability distribution over four variables $A_E, B, C_E, D$ which could be meaningfully compared to the predictions of quantum mechanics. 

So this  proposal delivers a kind of `indefiniteness' in Alice's outcome which is not in conflict with any of our ordinary beliefs about the absoluteness of observed outcomes. For example, suppose that at the time of the measurement, Alice  says to herself the words `I am currently seeing the result `up' relative to my axis.'  Then the story above affirms that at later times she will say to herself  `I remember seeing the result `up' relative to my axis' and in the case where Bob asks Alice about her outcome he will hear Alice say the words `I saw the result `up' relative to my axis,' and all of these words mean the same thing to everyone throughout the whole experiment, in an absolute sense; memory and communication  work in exactly the way we would ordinarily expect. So in this proposal, it is true in an absolute sense that if Bob hears Alice saying that she saw `up' relative to her axis then she  really did see `up' relative to her axis; it is simply the case that  only when she interacts with Bob in the right way does it become fixed what the direction `up relative to Alice's axis' corresponds to in \emph{Bob's} frame of reference, so her outcome is  indefinite and yet there is also an absolute fact about what she has observed.

\subsection{State Description \label{statedescription}}

As described above, the usual formulation of the Wigner's Friend and Extended Wigner's Friend scenarios supposes that when Alice performs her measurement her state relative to Bob transitions as follows: 

\[  \frac{1}{\sqrt{2}} (| \uparrow \rangle_S + | \downarrow \rangle_S) \otimes | \psi \rangle_A \  \rightarrow \ \frac{1}{\sqrt{2}}( | \uparrow \rangle_S | \uparrow \rangle_A + | \downarrow \rangle_S  | \downarrow \rangle_A ) \]

But the story above suggests a different state transition. In this proposal there is always a single definite fact about what Alice saw, perhaps selected at random, so let us consider the case where she observed the particle to be parallel to her axis. Then the overall state relative to Bob after the measurement should be represented as follows: 

 \[  \frac{1}{\sqrt{2}}  (| \uparrow \rangle_S | P \rangle_A  + | \downarrow \rangle_S | P' \rangle_A )  \]

Here, $| P \rangle_A $ represents a  state of Alice where she has observed the particle to be parallel to her own axis; $| P' \rangle_A $  also represents a state of Alice where she has observed the particle to be parallel  to her own axis, but it has a different total orientation relative to Bob.  That is,  $| P \rangle_A$ and $|P' \rangle_A$ both have the same \emph{internal} structure,  but they  stand in different relations to Bob. $|P \rangle_A$ and $|P' \rangle_A$ are identical from Alice's point of view as long as she has access only to facts about the inside of her lab, but they are not identical from Bob's point of view\footnote{ Another possible way of formalizing this proposal would be to treat `the facts about what Alice has seen' and `Alice's orientation with respect to Bob' as two distinct quantum degrees of freedom. Thus the post-measurement state relative to Bob, in the case where Alice gets the `parallel' outcome, would look something like$  \frac{1}{\sqrt{2}}  (| \uparrow \rangle_S | \uparrow \rangle_A | P \rangle_A  + | \downarrow \rangle_S  | \downarrow \rangle_A | P \rangle_A )$, where $| \uparrow \rangle_A$ and $| \downarrow \rangle_A$ correspond to different states for the orientation degree of freedom for Alice's laboratory, while $| P \rangle_A$ is the `parallel' state of the `what Alice has seen' degree of freedom. The reason I have not adopted this approach is because one possible interpretation of the proposal put forward here is that `what Alice has seen' should not be treated as a quantum degree of freedom at all, in which case it should not be represented as a quantum state in a Hilbert space.}. (If Alice had observed antiparallel instead, we would get a state of the same form with `parallel' replaced with `antiparallel'). 

Note that this state  is still a superposition relative to Bob in the orientation basis, so if he performs an appropriate measurement he will be able to detect interference between the branches, just as ordinary quantum mechanics predicts. That is, the results of probing this state in the orientation basis will be no different from the results of probing the state $\frac{1}{\sqrt{2}}( | \uparrow \rangle_S | \uparrow \rangle_A + | \downarrow \rangle_S  | \downarrow \rangle_A )$ as in the standard presentation of the Wigner's Friend experiment. But also, in both branches of the superposition Alice's internal structure encodes the result `parallel,' so when Bob performs a measurement to ascertain the \emph{internal} structure of her state (for example,  by simply asking her about her outcome) he will always learn that she saw parallel, regardless of whether or not her axis was found to be up or down relative to him. So we see that it is possible both for Alice and her laboratory to be in a superposition relative to Bob, and also for there to be a well-defined and communicable fact about what she has personally observed.

By analogy to the quantum reference formalism, we can also switch into Alice's reference frame and describe the situation from her point of view.  In the case where Alice saw $S$ to be parallel with her, of course the state of $S$ relative to her is $| \uparrow \rangle_S$. Meanwhile, since the orientation of $S$ is now well-defined relative to Alice but the relative orientation of Bob and $S$ is undefined, we can see that the orientation of Bob relative to Alice must now be indefinite. However, Bob still has a unique well-defined \emph{internal} configuration, which I will call $W$ -  the state of simply waiting while Alice performs her measurement. So the overall state relative to Alice will include two branches in which Bob has states $|W \rangle$ and $| W' \rangle$, such that these states are subjectively identical from Bob's point of view but they have different total orientations relative to Alice. Thus the state relative to Alice looks something like this: 

 \[  \frac{1}{\sqrt{2}}   | \uparrow \rangle_S \otimes (| W \rangle_B  + | W' \rangle_B  )  \]

We can straightforwardly apply this approach to the Extended Wigner's Friend scenario of \cite{Bong_2020}. Here for simplicity I will again assume that the relevant measurements are measurements of spin relative to some axis held in the laboratory, and I will take it that Alice and Chidi are measuring a pair of particles $X$, $Y$ originally prepared in the state $ \frac{1}{\sqrt{2}} (  | \uparrow \uparrow \rangle_{XY}     + | \downarrow \downarrow \rangle_{XY}   ) $. After Alice and Chidi have performed their measurements each of  them has a well-defined internal configuration which records a specific outcome of the  measurement, but they are still in a superposition in the orientation basis relative to the external frame occupied by Bob and Divya. Suppose for example that Alice saw `parallel' and Chidi saw `antiparallel.' Then the appropriate state is: 

\[ \frac{1}{\sqrt{2}} (  | \uparrow \uparrow \rangle_{XY} |P \rangle_A | A \rangle_C   + | \downarrow \downarrow \rangle_{XY} |P' \rangle_A | A' \rangle_C   )  \]

As in the previous case, $| P \rangle_A $ represents a state of Alice where she has observed the particle to be parallel  to her own axis and and $| P' \rangle_A $ is a subjectively identical state with a different overall orientation; similarly for  $| A \rangle_C $ and $| A' \rangle_A $.

Here it can be seen that in the orientation basis we have an ordinary quantum superposition relative to Bob and Divya, so if Bob and Divya perform their non-commuting measurements to obtain the values $B$ and $D$ they will see  results corresponding to the standard quantum predictions, as assumed in the Bong et al theorem. But if Bob instead tries to find out the value $A$, there are two different measurements that must be disambiguated. First, Bob could perform a measurement which probes the internal structure of Alice's laboratory, in which case he will find out that she saw parallel, corresponding to the variable $A_I$ as defined above. This variable has a definite well-defined value on every run of the experiment, but it will not exhibit the usual quantum correlations with $C$ or $D$ since it is the same on both branches of the superposition. Or alternatively Bob could perform a measurement which jointly probes Alice's outcome and also the direction of her axis relative to his own laboratory, corresponding to the variable $A_E$ as defined above. In this case he is in effect simply performing a measurement on the system $X$ in the superposition state  $(  | \uparrow \uparrow \rangle_{XY}  + | \downarrow \downarrow \rangle_{XY}  )$, and we can see that this experiment will reproduce the expected quantum statistics, although the result will look somewhat retrocausal since the value of $A_E$ as witnessed by Bob only becomes fixed when he actually decides to find out the value of $A$, and the same goes for Divya if she tries to find out the value $C_E$. 

Thus the state shown above is one in which the observed outcomes are absolute and yet it also reproduces the standard quantum predictions for measurements in the orientation basis, \emph{without} retrocausality or superdeterminism\footnote{Some non-locality will still be involved, but it is just the standard kind of Bell non-locality as we would expect for a measurement on the state  $(  | \uparrow \uparrow \rangle_{XY}  + | \downarrow \downarrow \rangle_{XY}  )$}.  So this approach seems to offer an appealing way of solving the problem posed by the EWF scenarios without sacrificing anything  essential in the process.

\section{Reference Frames and Properties \label{symmetry}}

In section \ref{Wigner} we considered the hypothesis  that at the time of Alice's measurement there is an absolute fact about what Alice has observed, but there is not an absolute fact about the orientation of Alice's reference frame relative to Bob's. The reason it is possible to do this is because `orientation' is linked with a symmetry of the laboratory subsystem. 

A theory is said to have a symmetry if there is a class of transformations $\{ T\}$ which leave its equations of motion and solutions unchanged, meaning that the effect of such transformations cannot be detected. For example, in most modern spacetime theories `global translations' and `global rotations,' where we displace or rotate the whole contents of the universe in the same way, are symmetries. It is natural to say that these transformations do not produce physically real change, and it follows that any variable $V$ whose value can be changed by these global transformations is not physically real. 

But we can also imagine applying transformations to proper subsystems of the universe. Suppose that $\{T \}$ is a symmetry group of the relevant theory, such that applying a transformation in $\{ T \}$ to a proper subsystem doesn't change the equations of motion or solutions for that subsystem. Thus applying a transformation in this  group to a subsystem is undetectable from the point of view of an observer within that subsystem - for example, if we rotate Alice's whole laboratory and its contents all together, this is not detectable from within the laboratory\footnote{I assume here that the laboratory is shielded in some way from the effects of gravity and other such external features which might make different orientations of the laboratory distinguishable; this is necessary because Bob will not be able to perform the supermeasurement to establish the value of $B$ if the laboratory is able to interact with the environment and undergo decoherence.}. However, the transformation may still be detectable when we compare the subsystem to its environment, since it may change the relation between the system and its environment in some way \citep{greaves2011empiricalconsequencessymmetries} - for example, rotating Alice's laboratory changes the relation of the laboratory to the external world, and clearly this is a physically meaningful difference since it will have real consequences if we open the door of the laboratory and Alice tries to exit.

So if there is some variable $V$ of a proper subsystem which is changed by the symmetry transformations $\{ T \}$, the value of $V$ is not meaningful when we consider the subsystem in isolation, but it can still be physically meaningful when compared to an external reference frame - for example  the orientation of Alice's laboratory is not meaningful for the laboratory in isolation, but it is physically meaningful when we describe the laboratory relative to the external world. Only variables which live in the invariant subspace of the subsystem with respect to the symmetry group $\{ T \}$ are physically meaningful in isolation, and normally we would imagine that it is those variables which would be be accessible to and experienced by an observer internal to the system - for example, presumably Alice's conscious experiences are defined by the internal configuration of the parts of the laboratory relative to each other and do not depend on the orientation of the laboratory as a whole relative to Bob. 

Thus it seems natural to say that in general a property $P$ of a subsytem needs to be relativized to a reference frame if and only if there is a relevant symmetry transformation for the subsystem that changes the value of $P$; for example, position must be relativized to a reference frame in any theory which has a translation symmetry, and  the orientation of spin must be relativized in any theory which has a rotation symmetry. The quantum reference frame research program in particular relies heavily on this way of thinking about relativization: `Note first that every reference frame is associated with a symmetry group. For instance, a Cartesian frame is associated with the group of rotations SU(2), a clock (phase reference) is associated with U(1), and a reference ordering (which we shall consider in section III F) is associated with the symmetric group SN 6. If Eve does not share Alice and Bob’s RF, then she is ignorant of which element of the group describes the relation between her local RF and that of Alice and Bob' \citep{PhysRevA.70.032307} That is, in this research programme it seems to be widely assumed that properties must be relativized if and only if they are linked in the right way with a symmetry of the theory. 

This suggests the following connection between definiteness and symmetries. Let us suppose that for any well-integrated subsystem $U$ such that all of its parts currently share a reference frame, any property $P$ of $U$ which is invariant under all relevant symmetry transformations will always have a well-defined value which is not relativized to any external system. But for a property $P'$ of the subsystem $U$  whose value can be changed by symmetry transformations performed on $U$, the property $P'$ has a well-defined value only relative to external systems. Thus there  may be some choice of external reference system $S$ such that the relation between $S$ and $U$ is indefinite, meaning that the value of the property $P'$ is indefinite relative to that system. Note however that this form of indefiniteness does not reflect any indefiniteness inherent to the subsystem $U$ itself;  only its relation with the external system $S$  is indefinite.

This seems like exactly the kind of distinction we need in order to resolve the EWF paradoxes. As above, consider the case where Alice is making a measurement of the orientation of spin. Since Alice, her laboratory and her axis together form a highly integrated system whose parts are constantly interacting and undergoing decoherence relative to one another,  they all belong to the same reference frame, and hence the supposition above suggests that the internal properties of this joint system which are invariant under symmetry transformations  must always be well-defined. Evidently this will include facts such as `Alice found the spin of the particle to be parallel to her axis,' and indeed it plausibly includes all of the facts on which Alice's conscious experience supervenes. But since Alice is isolated from the external world, her laboratory no longer belongs to the same reference frame as Bob, and thus since the rotation group is a symmetry of the laboratory subsystem, we can imagine that  after Alice performs a measurement on the orientation of the spin system, the resulting state can be written as a superposition such that in each branch of the superposition Alice's laboratory as a whole has a different orientation with respect to Bob's laboratory, but nonetheless the variables of her laboratory which live in the invariant subspace of the rotation group are the same in both branches. So we can maintain that there is a definite fact about what Alice observed and she has a single definite conscious experience, even though her laboratory is in a superposition of two different orientations relative to Bob.

 For illustrative purposes I have focused here on the case where the state of the system $S$ is a superposition of different orientations of spin, but similar stories can plausibly be made to work for other types of preparations. For example, suppose we instead prepare a state for the system $S$ which is a position basis superposition, such that in one branch the particle goes down the `left' path and in another branch the particle goes down the `right' path, with both paths initially defined relative to the external reference frame occupied by Bob. Let us suppose that at the time when Alice is still aligned with the external frame, she  puts a `red' detector into the left path and a `blue' detector into the right path; then in accordance with the account above, suppose that on any given run, the particle is detected in some particular detector - e.g. let us say that it is found in the red detector. But ex hypothesi when Alice performs this measurement she and her detectors lose their alignment with the external spatial reference frame, and thus since `left' and `right' were defined relative to the external frame, there is no longer a well-defined fact of the matter about whether the red detector is in the left path or the right path relative to the external frame.  That is, from Bob's point of view Alice is now in a superposition of one branch in which the red detector is on the left path, and another branch in which the red detector is on the right path, i.e. these two states correspond to different overall positions of Alice's whole laboratory relative to Bob's laboratory:

 \[  \frac{1}{\sqrt{2}}  (| \uparrow \rangle_S | R \rangle_A  + | \downarrow \rangle_S | R' \rangle_A )  \]

Here, $|R \rangle$ and $| R' \rangle$ both correspond to states in which the particle is found in the red detector, but from Bob's point of view they differ by a spatial translation. So as in the spin case, the state shown above is a superposition relative to Bob but nonetheless it represents a definite fact about the outcome that Alice saw and her corresponding conscious experiences. 
 
  Similarly, suppose we instead prepare a state for the system $S$ which is an energy basis superposition, i.e. $S$ has different energy levels in the two branches. Since energy is meaningful only relative to a standard of zero energy, there is plausibly a symmetry corresponding to changing the zero of energy, so  we can imagine that this results in a superposition of two states of Alice's laboratory which internally look the same but which differ in their overall energy with respect to the standard of zero energy associated with Bob's external reference frame. Thus again we can get a state which is a superposition relative to Bob but which represents a definite fact about the outcome Alice saw and the conscious experiences she is having. 

Evidently in order to make this approach work for all conceivable versions of the EWF experiment, one would need to hypothesize that something similar works for \emph{all} quantum degrees of freedom. That is, any property $P$ of a subsystem $U$ which can be superposed relative to an external reference frame is associated with a symmetry transformation which changes the value of $P$ but which leaves the equations of motion and solutions for $U$ as a whole  unchanged. I think it is plausible that this is indeed the case, but I will return to the question in section \ref{regress}.
 
In this article I will not speculate too much on the details of a mechanism which might underlie the forms of definiteness and indefiniteness suggested above. However, one natural proposal is to simply say that interactions between systems always happen in a single definite way. Thus for an integrated subsystem $U$ whose parts are in continuous interaction as they undergo decoherence relative to one another, the internal properties of $U$ which play a role in these definite interactions - i.e. relations between parts of $U$ which are invariant under all relevant symmetry transformations - are necessarily well-defined at all times.  But by definition a property $P'$ of $U$ as a whole which is changed by symmetry transformations cannot play any role in these internal interactions, so the value of $P'$ relative to an external system $S$ with which $U$ is not currently interacting does not need to be well-defined. Thus this supposition provides a principled way to distinguish between well-defined properties internal to the parts of a subsystem versus potentially indefinite properties relative to non-interacting external systems. We may then perhaps postulate that conscious experiences supervene  on interactions rather than on `states,' so this supposition would guarantee that we never end up with superpositions of different conscious states, thus resolving the Wigner's Friend scenario in a way which does not require us to postulate any kind of indefiniteness of conscious experiences. 

\subsection{Indefinite relations \label{physical}}

I have suggested that in an Extended Wigner's Friend experiment, a property such as orientation of Alice's laboratory relative to Bob's laboratory can sometimes be indefinite, or equivalently some relation between the laboratories such as relative orientation can be indefinite. But what kind of indefiniteness is required here? 

One might first think that this could be merely  a kind of `epistemic' indefiniteness, such that Alice's laboratory always stands in a well-defined relation of this kind relative to Bob's laboratory, but since Bob does not know what that relation is there are some limitations on what he can achieve operationally, similar to the kinds of limitations studied in the quantum information appraoch to quantum reference frames \citep{2021qrft,Bartlett_2007,PhysRevA.99.052315}. Can such an epistemic view give us what we need to resolve the EWF paradoxes? 

As a first pass, suppose we posit that Alice always has the \emph{same} well-defined orientation relative to Bob. But in this case, in the course of many trials Bob could figure out what the unknown orientation is, either by asking Alice directly during the interaction in which he acquires the value of $A$, or by comparing the results of the measurements to the quantum predictions and doing some calculations. And then he could simply instruct Alice to change the direction in which she measures her particle relative to her own laboratory reference frame in such a way as to compensate for their relative orientation, thus reproducing the EWF experiment as it was originally designed. So in this case the contradiction would not be resolved. 

Alternatively, suppose we posit that on every run of the experiment Alice's system takes on a \emph{different} unknown  orientation relative to Bob's frame of reference, perhaps at random. That is, Alice's axis always has some well-defined orientation relative to Bob's system, but it cannot be known in advance what that orientation is, and subsequently the only possible way to find out what it is involves Bob performing the measurement to learn $A$ rather than $B$.  In particular, this means that  when Alice and Bob talk after the experiment, they will find that Alice's axis does not always become aligned with Bob's axis in the same way: on some runs of the experiment   Alice's axis is parallel to Bob's, and on other runs of the experiment   Alice's axis is antiparallel to Bob, so some kind of random choice is being made when Bob performs this measurement.

However, in fact this approach will not work either. For if Alice's axis always has a well-defined orientation relative to Bob's, then the variable $A_R$ is in fact well-defined on every run, even if Bob does not know its value, and thus there must exist a well-defined joint probability distribution over $A_I, A_R$ after all; likewise for $C_I, C_R$. Thus if we don't allow retrocausality then in this case it would seem to follow that then there must always be a well-defined probability distribution over $A_I, B, C_I, D, A_R, C_R$, and this implies that there is  a well-defined distribution over $A_E, B, C_E, D$, so the contradiction would still not be resolved.

Thus it can be seen that a purely epistemic form of indefiniteness will not work here:  in order to make the proposed solution work we must say that once Alice has performed her measurement there is simply \emph{no fact} about the orientation of Alice's laboratory relative to Bob's laboratory, unless and until Bob performs the measurement to acquire the value of $A$. At that point, and not before, we get something like a quantum jump in which Alice's axis randomly `chooses' some orientation in Bob's reference frame; if Bob does not acquire the value of $A$, Alice's axis never takes on a well-defined orientation relative to his frame and hence no well-defined probability distribution exists.

 \subsection{Maintaining Alignment}

 At this juncture one might raise the following worry. Of course Alice and Bob would have had a well-defined orientation relative to each other prior to the experiment, and it does not seem as though anything disruptive happens during the experiment to alter that - after all, in order to be able to do this experiment it is essential that Alice and her system should be well-isolated from the external world in order to avoid decoherence, and if nothing interacts with Alice to disturb her, one might naturally expect that she would retain her well-defined orientation relative to Bob indefinitely. 

There are two possible ways of answering this question. The first is to say that simply the act of completely isolating Alice from the external environment automatically makes her lose her alignment with that environment.  However, one might worry that this proposal is inconsistent with empirical evidence, since we know that at least for small quantum systems we can put them in a pure state of well-defined alignment relative to a laboratory frame and then isolate them and, at least in principle, they will remain in the same pure state relative to the laboratory for apparently arbitrary lengths of time. 

The alternative is to say that Alice can in general keep her well-defined orientation relative to Bob until she performs her measurement on the qubit $S$. Indeed, this makes sense: by definition $S$ does not have a definite orientation relative to Bob's reference frame, and in the process of performing the measurement Alice aligns her own laboratory with $S$, so if we maintain that Alice does in fact always see a single definite outcome of her measurement, then in the course of performing the measurement she \emph{must} cease to have a definite orientation relative to Bob's reference frame. 

But on reflection one might worry that this involves a puzzling  asymmetry. Why should the interaction between $S$ and Alice result in Alice's reference frame going out of alignment with Bob's reference frame, rather than $S$ being dragged into alignment with Bob's reference frame? The latter might seem more natural, given that Alice's laboratory frame is so much larger than $S$. And if such a thing were to happen it would  simply be a collapse of the wavefunction from Bob's point of view, after which he would not be able to detect interference in the orientation basis. 

As a first response, note that we should not think of the alignment as involving any literal kind of jumping or dragging - in order to literally jump or be dragged Alice's laboratory would have to already have an orientation relative to $S$ that it could jump \emph{from}, and the point is precisely that it does not have any such thing. So the size difference may not be particularly relevant, since this is not the kind of literal physical interaction where inertia would be relevant. 

In addition, it should be kept in mind that this particular scenario may work differently from the cases we are familiar with, in virtue of the fact that Alice is completely  isolated from the external reference frame. Of course if Alice were situated within  the external reference frame in the ordinary way -  i.e. if she were entangled with it and constantly interacting with it by means of decoherence interactions as classical observers usually are -  then we know that when she measured the particle it would take on a well-defined direction relative to the reference frame rather than Alice going out of alignment with the reference frame, because empirically we  never find that a macroscopic observer's orientation relative to the rest of the macroscopic world suddenly changes when they measure the spin of an quantum particle. But since Alice is isolated from the external environment in the EWF scenario, we can perhaps postulate that in the absence of interactions which maintain her alignment with the external reference frame her  relation with it is unstable, so the orientation is lost when she aligns herself with another system which does not have a well-defined relation to the external frame. 

In any case, note that from the point of view of resolving the Wigner's Frame paradox we do not actually need to decide between Alice losing her alignment or the system $S$ becoming aligned. For if we accept the hypothesis that on any given run there is a definite fact about whether the spin is found to be parallel or antiparallel to Alice's laboratory, it immediately follows that Bob cannot keep the laboratory in a superposition unless he is able to make it the case that Alice ends up losing her alignment with the laboratory frame rather than the spin coming into alignment with the laboratory frame. It is possible that the choice between these two possibilities ultimately comes down to technical details of how the experiment is performed - Alice will lose her alignment only if  Bob has fine enough control to be able to maintain the superposition and perform the interference experiment. So if we assume  that it is possible to perform the EWF experiment as originally designed then it must be possible for Alice to lose her alignment as described, but the alternative is simply that the EWF experiment as designed cannot be performed even in principle - either way the apparent contradiction is resolved.

\section{Regress}

At this point, one might be worried that this solution only works because the experiment was set up in such a way that we can plausibly say that the two different states $| P \rangle$ and $| P' \rangle$ for the two different orientations of Alice's laboratory relative to Bob are subjectively identical from Alice's point of view; whereas it is surely possible to design set-ups where Alice necessarily ends up in subjectively different states in different branches of a superposition, and in that case the solution proposed here could not be applied. 

For example,  \cite{rovelli2024alicesciencefriendsrelational} proposes the following scenario in support of the idea that the relativization involved here must iterate. `Take the case that ... Friend interacts with S and measures S only if a quantum experiment gives him one outcome rather than another one. In Wigner’s account of the content of the room, there is a superposition between a state where S has a value for Friend, and a state where it doesn’t.'  Evidently we cannot say that a state in which Alice has performed a measurement is subjectively identical to a state in which she has not performed any such measurement, so one might worry that the proposed solution will not work in more complex scenarios like Rovelli's. 

However, let us analyse Rovelli's proposed scenario using the approach of section \ref{statedescription}. Suppose that $S$ is initially prepared in the state $\frac{1}{\sqrt{2}}(| \uparrow \rangle_S + | \downarrow \rangle_S) $ relative to Bob. Then after Alice's initial measurement we have one of two possible states relative to Bob, depending on whether she obtained the outcome parallel or antiparallel: 

\[   \frac{1}{\sqrt{2}} (| \uparrow \rangle_S | P \rangle_A  +  | \downarrow \rangle_S | P' \rangle_A  )  \]

\[    \frac{1}{\sqrt{2}} (  | \uparrow \rangle_S  | A \rangle_A   + | \downarrow \rangle_S  | A' \rangle_A )  \]

We then take it that in the case where Alice finds the particle  to be parallel to her axis she performs a second measurement on a new system $Y$ initially prepared in the ready state $| \phi \rangle_{Y} = \frac{1}{\sqrt{2}} (| \uparrow \rangle + | \downarrow \rangle)$ relative to Bob.  If $S$ and $Y$ are both parallel or both antiparallel to Bob's frame Alice will necessarily find $Y$ to be parallel to her axis, since it is aligned with $S$, and in the case where $S$ and $Y$ have different directions relative to Bob's frame Alice will necessarily find $Y$ to be antiparallel to her axis, since it is oriented oppositely from $S$. So we have three possible final states: 

\begin{align*}   \frac{1}{\sqrt{2}} (| \uparrow \rangle_S | \uparrow \rangle_{Y} | PP  \rangle_A + | \downarrow \rangle_S | \downarrow \rangle_{Y} | PP  ' \rangle_A    )  \end{align*}

\begin{align*}  \frac{1}{\sqrt{2}}  (| \uparrow \rangle_S  | \downarrow \rangle_{Y}  | PA \rangle_A  +  | \downarrow \rangle_S | \uparrow  \rangle_{Y}   | PA ' \rangle_A   )  \end{align*}

\begin{align*}   \frac{1}{\sqrt{2}}  ( | \uparrow \rangle_S  | \phi \rangle_{Y} | \sim M  \rangle_A  + | \downarrow \rangle_S | \phi  \rangle_{Y} | \sim M' \rangle_A  )  \end{align*}
 
Here, $|PP\rangle_A$ is the state of Alice where she finds both $S$ and $Y$ to be parallel to her axis, $| P A \rangle_A$ is the state of Alice where she first finds $S$ to be parallel to her axis and then later finds $Y$ to be antiparallel to her axis, and $| \sim M \rangle_A$ is the state of Alice where she finds $S$ to be antiparallel to her axis and thus does not perform a second measurement. $|PP'\rangle_A$, $| P A' \rangle_A$ and and  $| \sim M' \rangle_A $ respectively are the associated states which are subjectively identical for Alice but oriented differently relative to Bob.

Importantly, all three final states are still in a superposition relative to Bob in the orientation basis, so he will be able to detect interference between the two branches, just as predicted by standard quantum mechanics; in this sense, no violation of the  quantum predictions will be detected. But there is nonetheless  a well-defined fact about the result that Alice has obtained and about whether or not she has  performed a second measurement, because the facts about whether or not she has performed a second measurement belong to the invariant subspace with respect to the rotation group. Thus when Bob talks to Alice he will always find that if she reports finding the $S$ particle parallel to her axis, then she also reports having performed the second measurement; but Alice finding the $S$ particle parallel to her axis does not necessarily correspond to the $S$ particle pointing `up' in Bob's own reference frame, since her axis is not always pointing along his `up' direction. That is, from Bob's point of view he will find that sometimes the second measurement is performed when the spin points up and sometimes it is performed when the spin points down, but there is always consistency between what Alice reports seeing and whether the second measurement was in fact performed.   

Why should we think this is the correct representation of the situation, as opposed to the one suggested by Rovelli? Other than simply the fact that it allows us to resist falling into a regress, a further reason is that it is not a given that a fact such as `S has a value relative to Friend' is the kind of property which can be superposed at all. There exist well-understood quantum operators for properties such as position and spin, but we do not have an explicit construction for an operator for `the fact that S has a value relative to Friend,' so we cannot explicitly construct the superposition that Rovelli posits except in loose, schematic terms\footnote{Thanks to Timotheus Riedel for this point.}. Of course, it is quite common within quantum foundations to simply appeal to linearity in order to conclude that certain kinds of macroscopic properties can be superposed, but perhaps this assumption deserves greater scrutiny. In particular, one might interpret the proposal here as an indication that quantum states \emph{just are} a tool for describing a scenario where a system does not stand in a well-defined relation relative to some reference frame, and then it is natural to conjecture that only quantities which fail to be invariant under symmetry transformations can be treated as quantum degrees of freedom and superposed; and since it is plausible that facts such as `S has a value relative to Friend' are invariant under all relevant symmetry transformations, perhaps they simply cannot be superposed.

\subsection{Adding More Systems}

One might nonetheless worry that the distinction made above collapses when  we bring in another external system. For simplicity, consider a theory with only the rotation group as a symmetry, and suppose that the orientation of $A$ relative to $B$ is indefinite. If we only consider descriptions of  $A$ relative to $B$ or $B$ relative to $A$ we can consistently maintain that all properties which are invariant under the rotation group always have definite values, but suppose we introduce another system $C$ and describe $A$ and $B$ together relative to $C$. In that case presumably the angle between $A$ and $B$ must be indefinite relative to $C$, and yet the angle between $A$ and $B$ is invariant under the rotation group, so we can no longer say that all properties invariant under the rotation group are definite. 

But it should be noted there are two different kinds of indefiniteness at play here. Given a system $A$, it does not have an orientation at all by itself, but we can always give it a well-defined orientation by adding an appropriate reference frame. So the indefiniteness of $A$ relative to $B$ is not a fact about $A$ at all but rather a reflection of the fact that $B$ is not an ideal reference frame to describe $A$. By contrast since the only relevant symmetry here is the rotation group, there is no reference system we can introduce which will make the angle between $A$ and $B$ well-defined. So the indefiniteness of this angle is quite different from the `top level' of indefiniteness associated with the orientation of individual systems, and thus there is no particular reason to think we are dealing with a regress: the angle between $A$ and $B$ is indefinite in an absolute sense which cannot be removed by further relativization, and thus there is no reason to add any further level of relativization.

In particular, recall that the hypothesis put forward in section \ref{symmetry} was that  for any \emph{integrated subsystem such that all of its parts share a reference frame}, properties of that subsystem which are invariant under the relevant symmetry transformations will always have a well-defined non-relativized value. That is, properties like `relative distance,' `relative orientation' which are invariant under the relevant symmetries  \emph{can} fail to be well-defined, but only between distinct non-interacting reference frames, not between parts of a subsystem continuously interacting within a given reference frame\footnote{It is possible that this distinction is related to the emergence of spacetime: for example, given the Page-Wootters formalism \citep{PhysRevD.27.2885,} and other similar relational approaches in quantum gravity it is plausible that we should think of spacetime itself as emerging relative to a choice of reference frame, in which case it seems quite natural that relative spatial properties are not always  well-defined \emph{between} reference frames - plausibly the relevant kinds of spatial symmetries are only meaningful in the first place within integrated systems which share a reference frame}. This hypothesis suffices to resolve the Wigner's Friend scenarios because consciousness is reliably associated with highly integrated interacting systems, so we can reasonably expect that for any conscious system there will always be enough absolute internal properties to determine a single well-defined conscious experience, even though various external relational properties may be indefinite. 

\subsection{Against Regress \label{regress}}
  
Clearly the putative resolution put forward here rests crucially on the idea that there is a well-defined distinction between properties which must be relativized to an external frame and properties which are meaningful without relativization. And here there is a potential problem, because some proponents of relational views of quantum mechanics have argued that \emph{all} properties must be relativized to an observer or reference frame, in which case the distinction used here would collapse. For example, \cite{rovelli2024alicesciencefriendsrelational} invokes the example described  above to conclude  that we get `an infinite regression that prevents us from making any meaningful absolute statement about what is the case.' 

However, the idea that  relativization of variables should be connected to physical symmetries can be used to furnish a quite general argument against such a regress. That is, suppose that as suggested in section \ref{symmetry} we take it that properties are relativized to reference frames if and only if they are associated with a corresponding physical symmetry. It follows that if we are going to posit a regress of relativization, then for every order  in the regress we would need to posit an additional symmetry group associated with that level of relativization. For example, for a theory which is invariant under global translations, positions can be derived from relative distances by choosing a reference system $S$ and stating positions relative to that frame. So in order to get a `regress of relativization'  we would need to find  some additional group of transformations $\{ T \}$ such that the equations of motions and solutions are  invariant under transformations in $\{ T \}$ but relative distances are changed by these transformations. In that case we could conclude that relative distances are also not physically real but must themselves be derived from some other quantity $Q$ by choosing a reference system $S$ and stating distances relative to $S$. 

Now, it does not seem impossible that there exists such a set of transformations  $\{ T \}$. For example, although our current best theories are not conformally invariant, one might conjecture that the best successor theory will turn out to be conformally invariant, in which case it would indeed be invariant under scale transformations which change the relative distances. And as long as there exists such a group, then presumably we can appeal to a general treatment such as that of \cite{de_la_Hamette_2020} to show how to move from descriptions in terms of the underlying quantity $Q$ to descriptions relative to a reference frame which specify relative distances. But  if there is truly a regress of relativization we would then need to posit yet \emph{another} symmetry group such that the theory is invariant under transformations in this group but the quantity $Q$ is changed by these transformations, and so on and so forth. That is, if there is really an infinite iteration of relativization then there would need to be an infinite number of symmetry groups, each removing some class of degrees of freedom from the theory. 

This seems problematic for two reasons. First, because there does not seem to be any empirical reason to believe that the kinds of symmetries  needed here actually exist. For example, consider Rovelli's proposal that `the fact itself that S has a value relative to Friend is only relative to something else' \citep{rovelli2024alicesciencefriendsrelational}.  Roughly speaking, if we suppose that Friend performed a measurement of orientation as in previous cases and obtained the outcome `up,' then `the fact that S has a value relative to Friend' corresponds to the fact that the particle is parallel with the rod used to define the Friend's axis. What kind of transformation makes the particle and the rod cease to be parallel but leaves all the dynamics and equations of motion unchanged? Global translations and global rotations will leave the particle and rod parallel, and so will diffeomorphisms, since angles are complete observables and hence are still physically meaningful in a diffeomorphism-invariant theory \citep{Rovelli_2002}. It is also hard to see how charge transformations, time transformations, parity transformations and so on could be relevant.  Of course we can always hypothesize that there are higher-level symmetries which haven't been discovered yet, but this would surely be quite speculative: we have a good understanding of the kinds of symmetry groups that realistic physical theories usually have, whereas it seems quite unmotivated to postulate an infinite number of unknown symmetry groups that we have no empirical reason to believe in. 
 
Second, if we allow that a given theory has the feature that all of its structure can be removed by appeal to an infinite series of symmetry groups, it is unclear in what sense this could be said to be a `theory' at all, since it would seem to assert precisely nothing. In particular, if we want our interpretation of quantum mechanics to maintain the claim that quantum mechanics is a correct description of some parts of the world at least at some effective empirical level, then we must surely leave at least enough structure to define Hilbert spaces and the Schr\"{o}odinger equation - for if even this structure is not physically real it seems hard to understand how it could possibly be correct to say that quantum mechanics is true or approximately true or a useful mode of reasoning in certain regimes, or anything of that kind. This suggests that there must be some baseline level of structure which is not removable, and thus we should resist the suggestion that there could be an infinite regress of relativization.

\subsection{Other Types of Relativization}

Now of course the proponents of the regress of relativization might respond to the prior argument by maintaing that the type of relavization that they are interested in is not in fact connected to symmetry groups but is some other kind of relativization, perhaps more metaphysical in nature. But there are problems with this kind of view. 

First, it looks quite uneconomical: we know that we do have various  quantities in physics which are necessarily relativized for the symmetry-based reasons stated above, and this kind of relativization can plausibly explain most of the `observer-relative' effects which have been noted in quantum contexts, so it seems redundant to add some other type of relativization on top of this. 

Second, it has often been argued that one major motivation for positing relativization in quantum mechanics is based on the fact that  relativization plays an important role in General Relativity. As \cite{vidotto2022relationalontologycontemporaryphysics} puts it, `The relationality that characterizes the relational interpretation of quantum mechanics is in fact not so unconventional after all. Rather, it characterizes modern physics ... relationality is present, perhaps in a transversal way, in virtually all aspects of contemporary physics.'  In particular, it seems to be hoped that this connection indicates that a relational view will be a helpful route toward unifying quantum mechanics and relativity. Yet if this is the motivation, then we should surely be seeking to use the same type of relativization in both quantum mechanics and relativity - and it is clear that the relativistic version \emph{is} associated to the type of relativization that comes from invariance under a symmetry group, i.e. in this case the diffeomorphism group, so we have good reasons to think that the quantum version should have a similar source.

Third, positing a regress of relativization necessarily means that we are declining to specify any concrete distinction between features of an experimental situation which are relativized and features which are not, since we are committed to saying that \emph{everything} is ultimately relativized.   But this is not a very sustainable approach, because in a relational view the predictions we make for a given situation depend sensitively on our decisions about what should be relativized and what should not be relativized. In the Wigner's Friend scenario this can be disregarded because we already know what results we expect to get and thus we can straightforwardly figure out what   choices we must make about relativization if we want to reproduce the ordinary quantum predictions - that is, we should relativize the quantum states and outcomes, but not the experimental setup or the structure of spacetime. But if we are ever to extend relational   approaches beyond the obvious paradigm cases into new settings and domains where we do \emph{not} know in advance what predictions we should expect to obtain, we will need  clearer criteria regarding what should be relativized - for example, this kind of issue  arises already when we try to extend relational approaches to various black hole paradoxes \citep{hausmann2025firewallparadoxwignersfriend}. On the other hand, appealing to the formalism of symmetry groups and reference frames seems to offer a well-defined and physically meaningful way of drawing this distinction, and thus we have good reasons to adopt such an approach rather than positing a  regress of relativization, in order that the approach can be formulated clearly enough to actually be  be applied to new cases.

\section{Healey's Pragmatism \label{intersubjectivity}}

 The  way of thinking about the relationship between quantum mechanics and reference frames suggests a potentially fruitful line of enquiry where we might understand the measurement problem as arising from a confusion between variables of a system which are only meaningful when relativized to an external system and variables of a system which do not need to be thus relativized. For example, we might posit that the quantum formalism should be understood a tool for describing the way in which systems acquire and then lose their alignment with a given reference frame, and thus only variables of the former type should be associated with quantum operators and subject to superposition.

In this approach quantum states would necessarily be relativized, as in various   relational and perspectival approaches to quantum mechanics
  \citep{Everett, 1996cr, brukner2015quantum, https://doi.org/10.48550/arxiv.1801.09307}. But in this picture quantum states are relativized not to individual observers but to `reference frames.' One reason this matters is because if descriptions are relativized directly to individual observers then we quickly run into  problems where  different observers end up trapped within their own sets of relative facts with no ability to access facts relativized to other observers \citep{https://doi.org/10.48550/arxiv.2203.16278,adlamnonabsolute}, as discussed in section \ref{Wigner}. Whereas if states are relativized to entire reference frames then presumably the same facts are accessible to all observers within the same reference frame, so we don't encounter the same kinds of epistemic problems. 

An example of such an approach is given by Healey's `desert pragmatism,' which involves  positing that quantum states and quantum measurement outcomes are relativized to an `agent-situation,' i.e. `a physically characterized situation that may (but need not) be occupied by a physically situated agent' \citep{Healey2012-HEAQTA}. It is plausible that agent-situations are similar to what I have called reference frames, particularly if we also suppose that in general `reference frames' exist on macroscopic scales\footnote{\cite{Healey2012-HEAQTA} states that he understands `agent-situations' to be different from reference frames - possibly because agent-situations need to be relatively large and macroscopic, whereas the literature on quantum reference frames suggests that even quite small physical systems can play the role of reference frames. However, when we focus on sufficiently macroscopic reference frames then it seems likely that the categories of agent-situation and reference frame will be fairly close together, since agents must certainly occupy something like a reference frame.}.  Moreover, \cite{pittphilsci20846} emphasizes the role of decoherence in producing and stabilizing these agent-situations, stipulating that relative facts arise `only in the perspective provided by a situation involving extensive environmental decoherence,' and this accords with the suggestion I have made that decoherence plays an essential role in maintaining the integration of systems within a reference frame.

However,  Healey's proposal is explicitly framed as a pragmatist approach, since Healey maintains that it is not intended to answer the question  ‘How could the world possibly be how this theory says it is?’ \citep{Healey2012-HEAQTA}. By contrast, I think that emphasizing the role of symmetries and reference frames suggests a route toward  telling a story about `how the world is' which fits naturally with the kind of approach that Healey endorses. For while the view I have suggested here hinges on the understanding that certain kinds of properties are properly understood as relational,   there is no need to relativize any `facts' in a particularly radical sense. As long as we resist the regress of relativization, it follows that for any integrated system possessed with some subspace which is invariant with respect to all relevant symmetries, there exist  facts about that system  which need not be relative to anything - i.e. facts about variables defined within its invariant subspace.  So we end up with a fairly  `realist' picture, where we can have well-defined non-relative facts about appropriate subsystems within macroscopic reference frames.

Note also that although it is natural in this picture to say that quantum states are relative to reference frames, they are not subjective and nor are they merely descriptions of perspectives or degrees of belief. Rather quantum states characterize a scenario in which the relation between a system and a reference frame is undefined, and they predict probabilities which characterize the actual relative frequencies with which such systems  will adopt   values for relevant variables when they come back into alignment with the reference frame. In particular, if we resist the regress of relativization then there will be an objective matter of fact about what these frequencies are, so the probabilities in question are `objective' and in that sense quantum states are also `objective' even though they describe relations between things.

\section{Conclusion}

We have seen that making use of the reference frame formalism to resolve the Wigner's Friend paradox has a number of advantages. First it allows us to do justice to the intuition that there is something `indefinite' about the measurement outcomes in this situation, but without isolating observers within islands of relative facts or sacrificing the reliability of communication, memory or language. Second, it provides us with the tools to resist the threat of an infinite regress of relativity, and potentially a concrete way of deciding what should and should not be relativized in novel applications of the relational view. Third, it offers a fruitful direction in which relational views of quantum mechanics could be developed, avoiding some of the intersubjectivity worries that plague traditional relational views. 

This way of thinking about Wigner's Friend and quantum superposition  seems to suggest some interesting possibilities for future work on the measurement problem, but of course it is not by itself a complete solution to the measurement problem. In particular, one of the main motivations for trying to solve the measurement problem has always been the idea that we should not  treat macroscopic observers as fundamental - i.e. as in the usual reductionist tradition we should be able to derive observers from something more fundamental. Whereas here I have simply presupposed macrosopic reference frames relative to which the quantum states are defined, thus in effect also presupposing macroscopic agents; so a full solution  would presumably require telling some story about how the reference frames emerge and what they emerge from.

In particular, if we take it that as proposed in section \ref{intersubjectivity} the purpose of the quantum formalism is to describe the way in which systems take on properties relative to reference frames, then presumably one could not hope to use the quantum formalism to describe the emergence of the reference frames themselves. It is also tempting to posit that the quantum formalism is really only appropriate to describe systems relative to   \emph{macroscopic} reference frames, given that all of the evidence we actually have available to us pertains to effects observed in macroscopic frames, and if this is so  one might worry that we cannot use quantum mechanics  to describe the microscopic world in the absence of a macroscopic reference frames. These considerations suggest that in order to describe the emergence of the reference frames we would need to move beyond quantum mechanics to some other kind of formalism, and this can perhaps be understood as providing an explanation for the fact that the measurement problem has been so intractable - perhaps it is simply not the kind of problem which we should expect to be able to solve within the quantum formalism itself.

 \section{Acknowledgements}

 Thanks to Anne-Catherine de la Hamette for very helpful comments on a draft of this paper, and to the participants of the Emerge Nordita workshop for helpful conversations. This work was supported by the  John Templeton Foundation Grant ID 63209.

\end{document}